\documentclass[12pt]{article}

\usepackage{scicite}

\usepackage{times}

\usepackage{graphicx}
\usepackage[closeFloats]{fltpage}

\newcounter{lastnote}

 \makeatother

\title{{\it Supporting Online Material for\/}\\From a single-band metal to a high-temperature \\superconductor via two thermal phase transitions}

\author
{Rui-Hua He,$^{1, 2, 3\ast}$ M. Hashimoto,$^{1, 2, 3\ast}$ H. Karapetyan,$^{1, 2}$ J. D. Koralek,$^{3, 4}$ \\J. P. Hinton,$^{3, 4}$ J. P. Testaud,$^{1, 2, 3}$ V. Nathan,$^{1, 2}$ Y. Yoshida,$^{5}$ Hong Yao,$^{1, 3, 4}$ \\K. Tanaka,$^{1, 2, 3, 6}$ W. Meevasana,$^{1, 2, 7}$ R. G. Moore,$^{1, 2}$ D. H. Lu,$^{1, 2}$ S.-K. Mo,$^{3}$\\M. Ishikado,$^{8}$ H. Eisaki,$^{5}$ Z. Hussain,$^{3}$ T. P. Devereaux,$^{1, 2\dag}$  \\S. A. Kivelson,$^{1\dag}$ J. Orenstein,$^{3, 4\dag}$ A. Kapitulnik,$^{1, 2\dag}$ Z.-X. Shen$^{1, 2\dag}$\\
\\
\normalsize{$^{1}$Geballe Laboratory for Advanced Materials, Departments of Physics}\\
\normalsize{and Applied Physics, Stanford University, Stanford, California 94305, USA}\\
\normalsize{$^{2}$Stanford Institute for Materials and Energy Sciences,}\\
\normalsize{SLAC National Accelerator Laboratory, Menlo Park, California 94025, USA}\\
\normalsize{$^{3}$Advanced Light Source and Materials Sciences Division,}\\
\normalsize{Lawrence Berkeley National Laboratory, Berkeley, California 94720, USA}\\
\normalsize{$^{4}$Department of Physics, University of California, Berkeley, CA 94720}\\
\normalsize{$^{5}$Nanoelectronics Research Institute, AIST, Ibaraki 305-8568, Japan}\\
\normalsize{$^{6}$Department of Physics, Osaka University, Toyonaka, Osaka 560-0043, Japan}\\
\normalsize{$^{7}$School of Physics, Suranaree University of Technology and}\\
\normalsize{Synchrotron Light Research Institute, Nakhon Ratchasima, 30000 Thailand}\\
\normalsize{$^{8}$Japan Atomic Energy Agency, Tokai, Ibaraki 319-1195, Japan}
\\
\normalsize{$^\ast$These authors contributed equally to this work.}\\
\normalsize{$^\dag$To whom correspondence should be addressed; E-mail: zxshen@stanford.edu,}\\ \normalsize{aharonk@stanford.edu, jworenstein@lbl.gov, kivelson@stanford.edu, tpd@stanford.edu.}
}

\date{\small{Submitted on 28 September 2010}}

\begin{document} 


\baselineskip24pt


\maketitle





\newpage
\tableofcontents
\vskip\baselineskip
\vskip\baselineskip
\vskip\baselineskip

\section{Materials and Methods}
\subsection{Samples}

\begin{figure}
\hspace*{-0.5cm}
\includegraphics [width=6.8in]{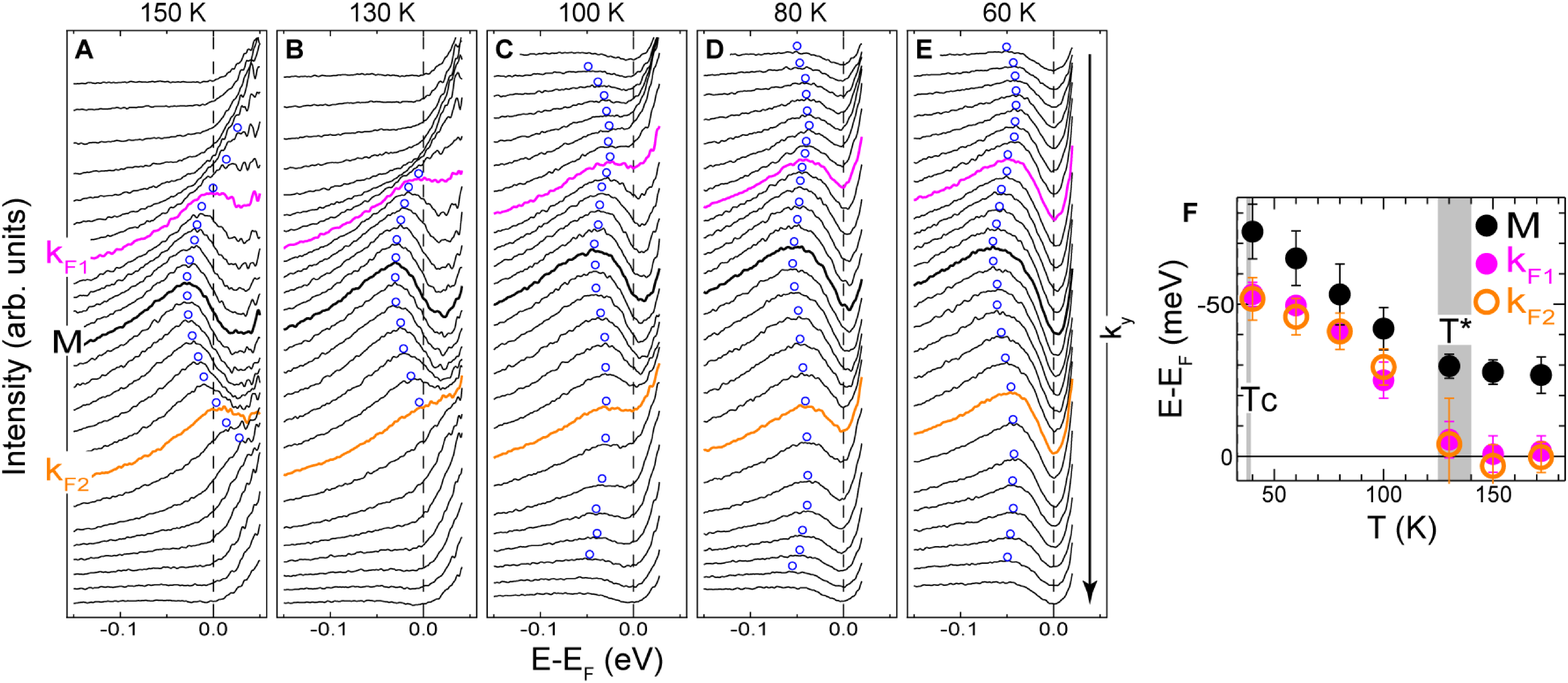}
\caption[ARPES temperature dependence across $T^*$] {\textbf{ARPES temperature dependence across $T^*$} \textbf{A}-\textbf{E}, Selected EDCs at five representative temperatures along Cut C1 shown in Fig. 1. The single EDC maximum is tracked to give the band dispersion summarized in the right inset of Fig. 3. \textbf{F}, Summary for the temperature-dependent binding energy position of the EDC maximum at M, $k_{F1}$ and $k_{F2}$. Error bars are estimated based on the sharpness of features.}
\label{Fig. S1}
\end{figure}

High-quality single crystals of Pb$_{0.55}$Bi$_{1.5}$Sr$_{1.6}$La$_{0.4}$CuO$_{6+\delta}$ (Pb-Bi2201) and\\ Bi$_{2-x}$Pb$_x$Sr$_2$CaCu$_2$O$_{8+\delta}$ (Pb-Bi2212) near optimal doping were grown by the travelling-solvent floating zone method. The Pb doping suppresses the super-modulation in the BiO plane and minimizes complications in the electronic structure due to photoelectron diffraction. The carrier concentrations of the samples were carefully adjusted by a post-annealing procedure under flowing $N_2$ gas which varies the oxygen content, and were estimated from the Fermi surface volumes, e.g., for Pb-Bi2201 shown in Fig. 1 to be $\sim$ 25.3 \%, which is consistent with previous reports for near optimal doping \cite{cuprates:Bi2201:KondoARPESvsResistivity, cuprates:Bi2201:MakotoBi2201dopingdep, HTSC:PseudogapDispersion}. X-ray and Laue diffraction showed no trace of impurity phases. For Pb-Bi2201, the onset temperature of the superconducting transition, $T_c$, determined by SQUID magnetometry, was 38 K, with a resistive transition width less than 3 K ($T_c= 97 \pm 2$ K for Bi2212 related to Fig. 4D); the pseudogap temperature, $T^*=132$ K $\pm 8$ K, is determined based on the observed closing of the antinodal pseudogap in ARPES (Fig. S\ref{Fig. S1}F), which coincides within experimental uncertainty with the onset of Kerr signal in PKE and that of transient reflectivity in TRR (Fig. 3). This $T^*$ value is also consistent with our resistivity measurement on the same crystals as well as other reports of different measurements on Bi2201 near optimal doping \cite{HTSC:PseudogapDispersion, cuprates:Bi2201:DingHongTwoGaps, HTSC:KondoCompetingPseudogapSC, cuprates:Bi2201:AndoTwoPseudogaps, cuprates:Bi2201:QCP1}. We found $T^*$ depends more sensitively on the post-annealing condition than $T_c$ near optimal doping but remains consistent among samples within the same growth and post-annealing batch.



\subsection{Measurements}
\subsubsection{ARPES}
Angle-resolved photoemission spectroscopy (ARPES) measurements were performed at Beamline 5-4 of the Stanford Synchrotron Radiation Lightsource (SSRL) with a SCIENTA R4000 electron analyzer. Preliminary photon energy and polarization dependent ARPES measurements were performed at Beamline 10.0.1 of the Advanced Light Source (ALS). All presented data were taken using 22.7 eV photons mainly in the first Brillouin zones with total energy and angular (momentum) resolutions of $\sim$ 10 meV and $\sim$ 0.25$^\circ$ ($\sim$ 0.0096\AA$^{-1}$), respectively. The temperatures were recorded closest to the sample surface position within an accuracy $\pm$ 2 K. The samples were cleaved and measured in an ultra high vacuum chamber with a base pressure of better than $3\times 10^{-11}$ Torr which was maintained below $5\times 10^{-11}$ Torr during the temperature cycling. Measurements on each sample were completed within 48 hours after cleaving. All data shown in the main text were obtained on the same sample and were reproduced on different samples from the same batch with sample cleaving at low (10 K) or high ($\sim$ 150 K) temperature. The possibility of sample aging in each experiment was excluded by monitoring the nodal dispersion without complications by the gap opening. A similar experimental routine has been previously demonstrated and detailed in the supplementary information of Ref. \cite{HTSC:PseudogapDispersion}.

\subsubsection{Polar Kerr effect}
Polar Kerr effect (PKE) measurements were performed using a zero-area-loop Sagnac interferometer at a wavelength of $\lambda=1550$ nm, spot size of $\sim 3\mu$m, and sensitivity of 0.1 $\mu$rad/$\sqrt{Hz}$ at 250 $\mu$W of incident optical power \cite{HTSC:KerrEffect_1}. In its used configuration, the apparatus was sensitive to only the polar Kerr effect, hence to any ferromagnetic component of the local magnetization, perpendicular to the plane of incidence of the light. The same apparatus was previously used to detect the effect of time-reversal symmetry breaking below $T_c$ in Sr$_2$RuO$_4$ \cite{HTSC:KerrEffect_SrRuO} and in the vicinity of $T^*$ in YBa$_2$Cu$_3$O$_{6+x}$ (YBCO) \cite{HTSC:KerrEffect_YBCO}. 

For YBCO, measurements on crystals with various doping levels revealed a sharp phase transition at a temperature $T_s$(p) below which there is a non-zero Kerr rotation. Both the magnitude and doping dependence of $T_s$ were found to be in close correspondence with $T^*$ which has been identified in other physical quantities. In particular, $T_s$ is substantially larger than $T_c$ in underdoped materials, but drops rapidly with increasing doping, so that it is smaller than $T_c$ in a near optimally-doped crystal and extrapolates to zero at a putative quantum critical point under the superconducting dome. The magnitude of the Kerr rotation in YBCO is smaller by $\sim$ 4 to 5 orders of magnitude than that observed in other itinerant ferromagnetic oxides, suggesting that at most it measures a very small ~``ferromagnetic-like" component of a magnetic transition. However, the temperature dependence of the PKE near $T_s$, together with complementary measurements using other probes, strongly suggest that the PKE tracks a secondary order parameter which is driven by another transition that is not necessarily magnetic. Corroborating evidence in that direction comes from neutron measurements on x= 0.45 \cite{stripe:neutron:YBCO_eLiquidCrystal} finding evidence for a nematic state in this composition, and from muon-spin-rotation measurements indicating charge ordering transition \cite{stripe:other:mSR_YBCO} near optimal doping. Interpolating between these two ends which find $T^*$ that agrees with the PKE measurements, we speculate that the PKE tracks an electronic transition that may evolve smoothly from strong charge ordering to a weaker nematic phase. 

While the above discussion on YBCO relies on circumstantial evidence, the present results on Pb-Bi2201 may be the first unambiguous data that suggests that $T^*$ can be identified, which corresponds to a true symmetry-breaking transition and the primary order parameter that governs that transition is electronic.

\subsubsection{Time-resolved reflectivity}
Time-resolved reflectivity (TRR) measurements were performed using a 100 MHz mode locked Ti: Sapphire oscillator operating at a wavelength of 800 nm. In these measurements, femtosecond (fs) pulses of linearly-polarized light excite the sample, and time (t)-delayed probe pulses from the same laser measure the resulting change in reflectivity, $\Delta R$. In the present context, the power of this technique lies in its ability to distinguish different phases by their response dynamics. Fig. S\ref{Fig. S3} shows TRR data taken on Pb-Bi2201 at 3 different pump powers with a focal spot diameter of roughly 100 $\mu$m. The grey and pink shaded regions highlight the pseudogap and superconducting responses, respectively. As discussed in the main text, the superconducting signal, seen in Fig. S\ref{Fig. S3}B \& C, is attributed to the breaking of Cooper pairs, and its temperature and pump-fluence dependence are consistent with previous work \cite{HTSC:Pumpprobe_pairing}. It is characterized by a positive signal which decays on the picosecond (ps) time scale. The pseudogap signal, on the other hand, is centered about zero delay and decays on the 100 fs timescale. The pseudogap signal is generally weaker than the superconducting one and becomes more pronounced with increasing pump power. However, the average heating of the sample becomes significant at low temperatures for high pump powers, which can almost completely suppress the superconducting response. The magnitude of the negative TRR signal in the left inset of Fig. 3 is extracted from the 15 mW data (Fig. S\ref{Fig. S3}A) by taking the difference between the maximum for $t< 0$ and the minimum near $t= 0$. Because of the laser heating at this power, we only show the data for $T\ge 25$ K, and estimate an error bar of $\pm 8$K for the actual sample temperature.

\subsection{ARPES data analysis}

To remove the effect due to cutoff by the Fermi-Dirac (FD) function, raw ARPES spectra have been divided by a convolution of the FD function at the given temperature and a Gaussian with its width fixed at the energy resolution. Note that this division is an approximation for the complicated deconvolution that aims at removing the energy resolution effect. It works well for high temperatures when the overall width of FD function ($\sim 4k_BT$) is much larger than the energy resolution and becomes bad for low temperatures. This procedure allows us to recover the actual band dispersions closest to $E_F$ and trace them above $E_F$ at high temperatures, when thermal population leads to appreciable spectral weight. It is also necessary for revealing that the EDC shoulder feature at low temperatures is not trivially produced by the Fermi cutoff. Because the uncertainty introduced by the division to the spectral line shape is limited to $\sim 4$ meV ($=4k_BT$) at 10 K around $E_F$, the discussed low-temperature spectral evolution across $T_c$ for features (such as the EDC shoulder feature) located beyond this $E_F$ vicinity remains robust. The faithfulness of this procedure can be reflected by the observations after the division that the EDC shoulder feature at low temperatures does not show up everywhere but is confined in the antinodal region and that the nodal Fermi crossings are recovered at all temperatures. However, one should exercise caution when associating the gap function at low temperatures directly with the peak position of the FD-divided spectrum (Fig. S\ref{Fig. S4}), particularly in the vicinity of the node where the gap is intrinsically small.

\begin{figure}
\hspace*{1cm}
\includegraphics [width=5.29in]{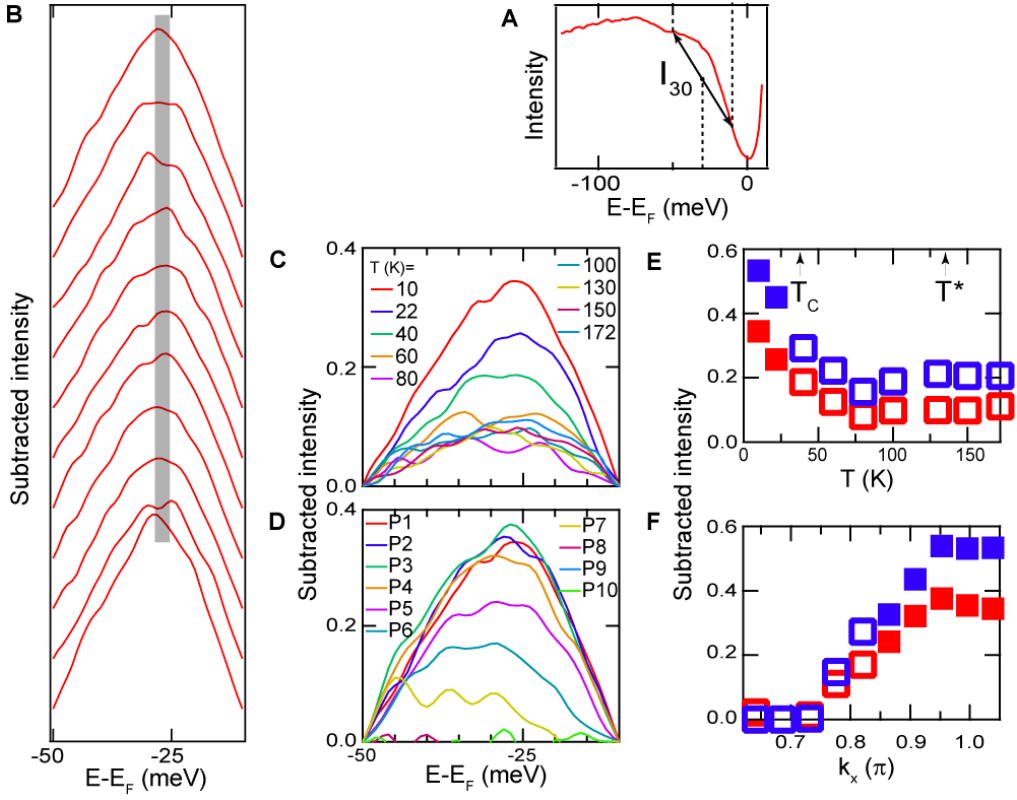}
\caption[Determination of the shoulder feature] {\textbf{Determination of the EDC shoulder feature by spectral subtraction.} Subtraction was made for each EDC (e.g., the EDC at M at 10 K of Cut C1 as shown in \textbf{A}) by assuming a linear background connecting intensity at E$_{i1}$=-50 meV and E$_{i2}$=-10 meV. Intensity was then normalized by the linear intensity (I$_{30}$) at E$_{i}$=-30 meV. The subtracted EDCs show a clear intensity maximum from which we define the position of the original EDC shoulder. This method yields a similarly non-dispersive peak at a similar energy position as the spectral division using the 40 K data (e.g., as shown in \textbf{B} for the same data set as in Fig. 4C). It also provides a convenient way to quantify the temperature- and momentum-dependent existence of the EDC shoulder feature. \textbf{C} shows the subtracted EDCs (\textbf{A}) at various temperatures and \textbf{E} plots the temperature dependence of the maximum intensity (red) in comparison to a different subtraction using E$_{i1}$=-60 meV and E$_{i2}$=0 meV (blue). \textbf{D} \& \textbf{F} show the results for P1-P10 (in Fig. 2W). We defined phenomenologically the threshold for existence of the EDC shoulder feature at the half-drop positions of the maximum intensity w.r.t. the background (solid/empty symbols for existence/non-existence in \textbf{E} \& \textbf{F}). This consistently supports the eyeball estimate that the EDC shoulder feature loses its clear definition above $T_c$ and beyond the antinodal region (along $\Gamma$-M). Note that its existence perpendicular to $\Gamma$-M could in principle extend wider than where is currently identified by green circles (Fig. 2O-R), due to its potential mixing with the EDC maximum feature near back-bending. Also note that the non-vanishing variation of the maximum intensity above $T_c$ in \textbf{E} is consistent with the idea that superconducting fluctations persist above $T_c$ but still well below $T^*$, in nearly OP Bi2201 \cite{HTSC:NernstEffect_Ong_new} and other cuprates \cite{HTSC:Pseudogap_THz_Orenstein, HTSC:Pseudogap_STM_aboveTcQPI}.}
\label{Fig. S2}
\end{figure}

\begin{figure}
\includegraphics [width=6.5in]{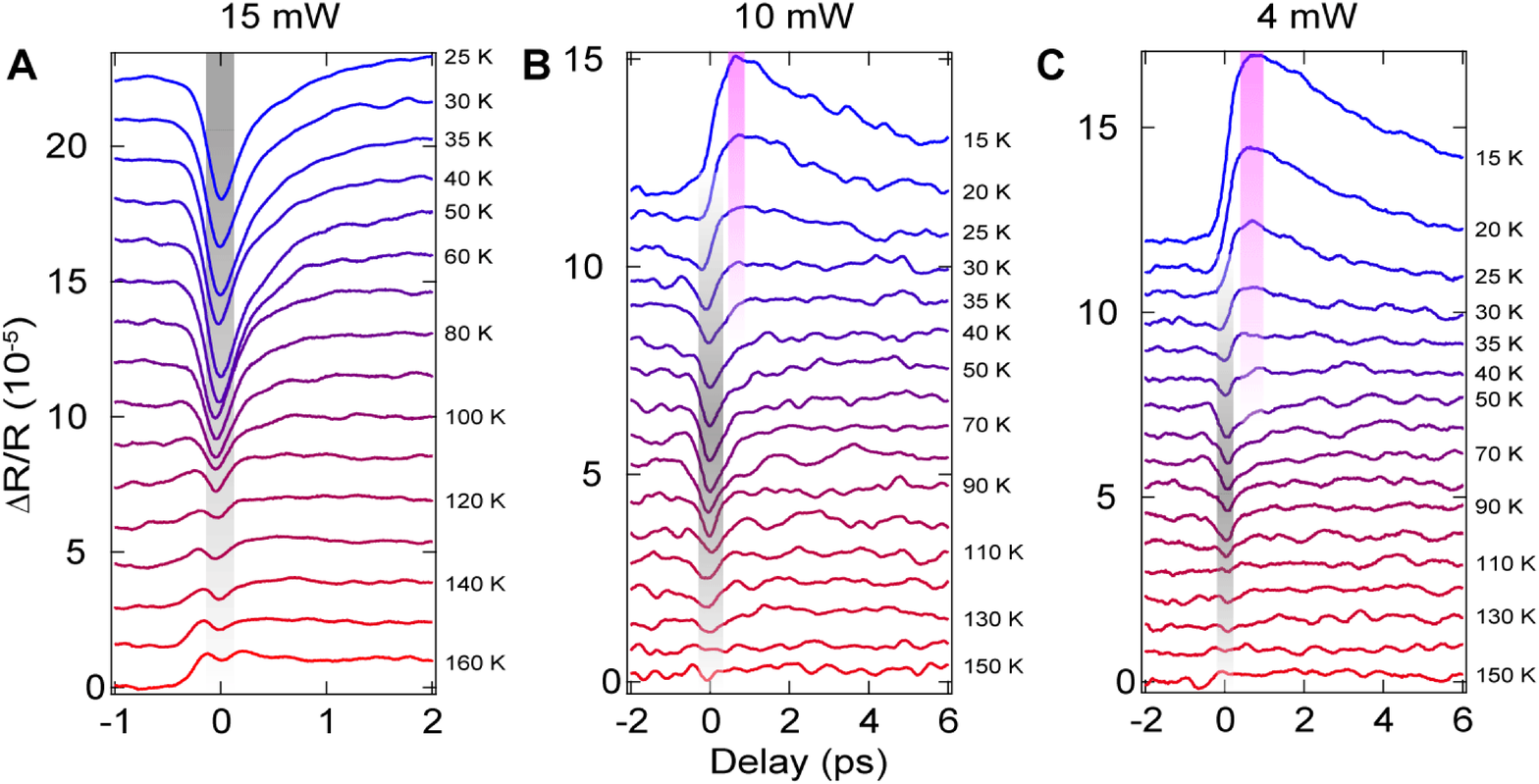}
\caption[TRR raw data] {\textbf{Temperature dependence of time-resolved reflectivity.} Results are shown at different laser powers, \textbf{A} 15 mW, \textbf{B} 10 mW, and \textbf{C} 4 mW. The grey and pink shaded regions highlight the pseudogap and superconducting responses respectively. Note that the 15 mW data was taken over a shorter delay range than the lower power data.} 
\label{Fig. S3}
\end{figure}

\begin{figure}
\hspace*{1cm}
\includegraphics [width=5.5in]{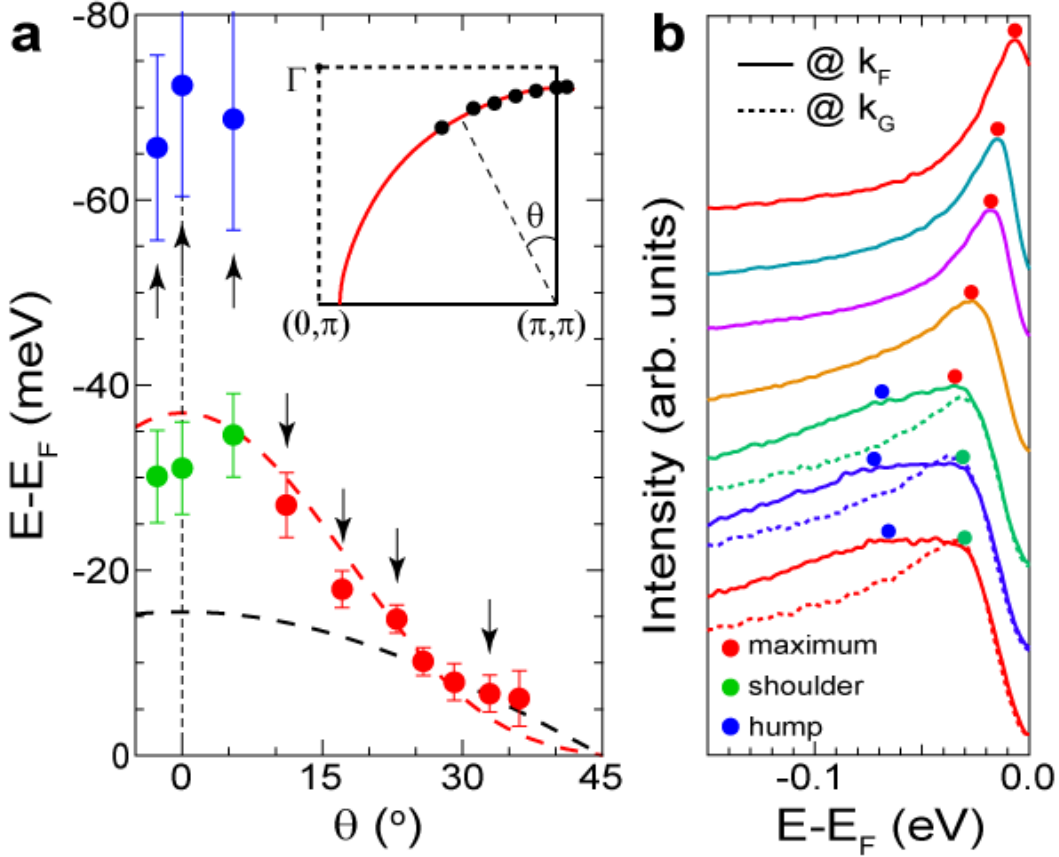}
\caption[Gap function] {\textbf{Gap function along the underlying Fermi surface at 10 K.} \textbf{A}, Binding energy positions of various spectral features in the EDCs at $k_F$ plotted as a function of the Fermi surface angle. Red dashed curve is the gap function reported for OP Pb-Bi2201 \cite{cuprates:Bi2201:Kondo_TwoGap}. Black dashed curve is a guide to the eye for a simple $d$-wave gap function. Our observation of multiple EDC features of comparable spectral weight in the antinodal region highlights the essential physics that eluded previous revelation. \textbf{B}, Selected EDCs at $k_F$ ($k_{F1}$) as indicated by black arrows (dots) in (the inset of) \textbf{A}. The maximum feature is defined for EDCs with a single component seen away from the antinodal region. Both the shoulder at low energy and the hump at high energy are defined for EDCs at $k_F$ in the antinodal region. The (antinodal) EDCs at the neighboring $k_G$ ($k_{G1}$) are shown in dashed curves, which have a single component, reminiscent of those used to define the gap function in the antinodal region in many previous studies.} 
\label{Fig. S4}
\end{figure}

The subsequent EDC analysis was performed on the FD-divided spectra: the EDC maximum below $E_F$ is identified by taking the first derivative of the moderately-smoothed EDC (w.r.t. the energy axis); suspected additional features below $E_F$ and local maxima above $E_F$ are estimated based on the first derivative and aided with eyeball correction; the (non-)existence and energy position of the EDC shoulder feature are determined mainly based on a spectral subtraction analysis (Fig. S\ref{Fig. S2}), which is in general consistence with the spectral division method using 40 K data (Fig. 4C) and with eyeball estimates.

\section{Simulations}
\subsection{Simple density-wave pseudogap order}
We describe modifications of the band structure due to $d$-wave superconductivity (Fig. S\ref{Fig. S7}B \& F) and its coexistence with some long-range density wave order of wave vector $q$ (Fig. S\ref{Fig. S7}C, D, G \& H) in a simple mean-field approach, modified from what we previously used in Ref. \cite{HTSC:PseudogapDispersion}. We confine the use of this approach only to low temperatures, where order parameter fluctuations are less severe and thus a simple mean-field description of the data might have a better chance of success. 

\begin{figure}
\hspace*{1cm}
\includegraphics [width=5in]{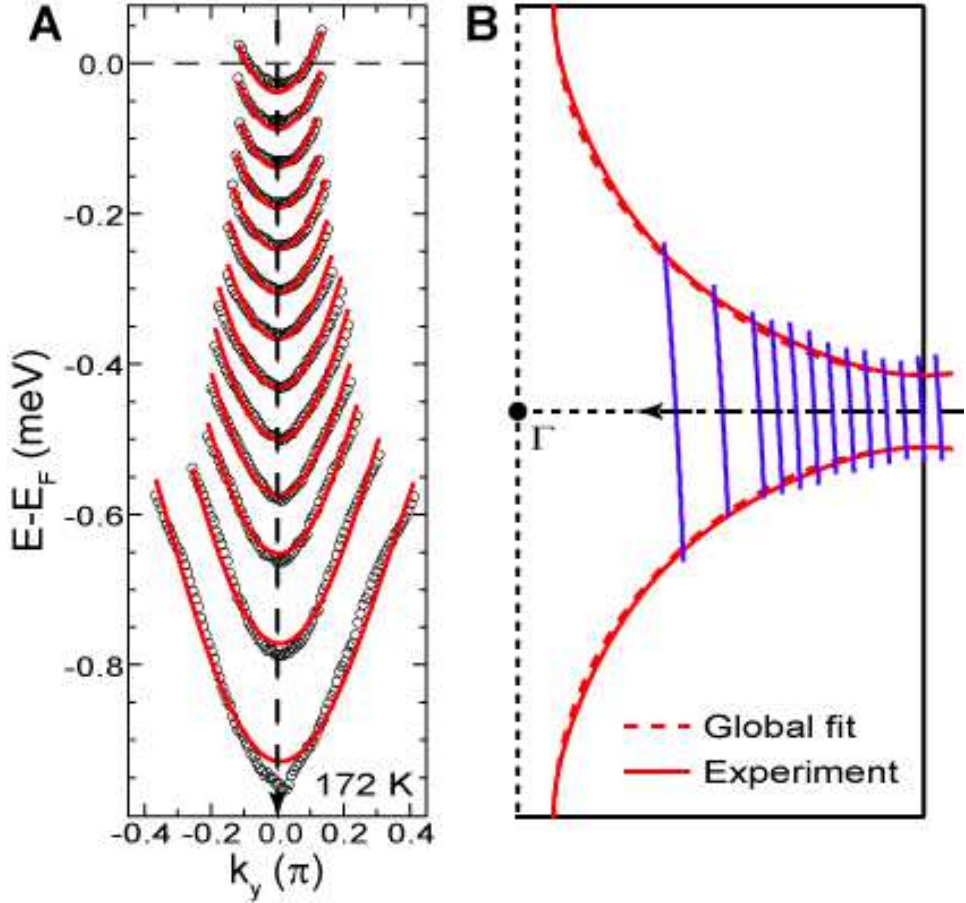}
\caption[Global tight-binding fit] {\textbf{Global tight-binding fit.} \textbf{A}, Global tight-binding fit (red curves) to the EDC dispersions at 172 K (black circles) whose momentum loci are indicated by blue lines in \textbf{B}. For clarity, adjacent cut has a 50 meV relative vertical offset. Because of the global constraint, the fit gives a slightly larger ($\sim 7$ meV) band bottom at M point than the experiment. \textbf{B} shows that the Fermi surface given by the global fit coincides with the experimental one reproduced from Fig. 1. Please note that the experimental band dispersions are obtained (throughout this work) based on the EDC analysis, which is the traditional way of extracting dispersion relations from photoemission; it also suffers much less from the momentum-dependent matrix elements than the momentum-distribution-curve analysis, such that a parabola-like dispersion, rather than the ~``waterfall" type, remains to be seen close to the $\Gamma$ point.}
\label{Fig. S5}
\end{figure}

The mean-field coexistence Hamiltonian is given by: $H = \sum_k\epsilon_kc^\dagger_kc_k+\sum_{k,q}V_q(c^\dagger_{k+q}c_k+h.c.)
+\sum_k\Delta_k(c^\dagger_{k\uparrow}c^\dagger_{-k\downarrow}+h.c.),$ where $V_q$ and $\Delta_k$ are the density wave and superconducting order parameters, respectively; $c_k^{\dag} (c_k)$ is the creation (annihilation) operators for electrons at $k$. $\varepsilon_k=-2t(\cos k_x+\cos k_y)-4t^\prime\cos k_x\cos k_y-2t^{\prime\prime}(\cos 2k_x+\cos 2k_y)-4t^{\prime\prime\prime}(\cos 2k_x\cos k_y+\cos k_x\cos 2k_y)-\varepsilon_0,$ where $t, t^\prime, t^{\prime\prime}, t^{\prime\prime\prime}, \varepsilon_0=0.22, -0.034315, 0.035977,-0.0071637, -0.24327$ eV, respectively, is the tight-binding bare band dispersion which is obtained by a global fit to the experimental EDC dispersions at 172 K (Fig. S\ref{Fig. S5}). 

We consider in the following only the density wave orders with generic $q$s that largely connect the antinodal portion (rather than the nodal one) of the Fermi surface, where the particle-hole asymmetry of spectra is the strongest. 

In case of bond-diagonal density wave order with a commensurate $q = q_{AF} = [\pi,\pi]$, the eigenstate $|\psi_k\rangle=u_k|k\rangle+u_{k+q_{AF}}|k+q_{AF}\rangle+u_{-k}|-k\rangle+
u_{-k-q_{AF}}|-k-q_{AF}\rangle$ for the Hamiltonian with the eigenenergy $\varepsilon^\prime(k)$ can be obtained by solving the matrix $M_{AF+SC}=\left[ \begin{array}{cc} A_{AF} & D_{SC}\\ D_{SC} & -A_{AF} \end{array} \right],$ where $A_{AF}=\left[ \begin{array}{cc} \varepsilon_k & V_2\\ V_2 & \varepsilon_{k+q_{AF}} \end{array} \right], D_{SC}=\Delta I, I$ is a $2\times 2$ identity matrix. We expect this case to generically represent density wave orders of $q$ approximately along the bond-diagonal direction, such as $d$-density wave, antiferromagnetic order with $q=q_{AF}$ and incommensurate spin stripes with a finite, small deviation of $q$ away from $q_{AF}$. 

In case of bond-direction density wave order, we consider the checkerboard case with $q_1 = [0.15\pi, 0]$ and $q_2 = [0, 0.15\pi]$, $M_{CB+SC}=\left[ \begin{array}{cc} A_{CB} & D_{SC}\\ D_{SC} & -A_{CB} \end{array} \right],$ where $D_{SC}=\Delta I, A_{CB} (I)$ are all $1600\times 1600$ (identity) matrices. $A_{CB}$ is determined by the following, base: $\{k+mq_1+nq_2; m, n=0,...,N-1; N=200/gcd(2*100,0.15*100)=40\}$; matrix element: $A_{CB}(m,n;m',n')=V_1$, only if $|m-m'|+|n-n'|=1$ or $(|m-m'|+|n-n'-N+1|)(|m-m'-N+1|+|n-n'|)=0$ [with small high-order interaction neglected \cite{CDW:VoitIncommDW}], and 0 otherwise. Eigenstate is thus $|\psi_k\rangle=\sum_{m,n}u_{k+mq_1+nq_2}|k+mq_1+nq_2\rangle+u_{-k-mq_1-nq_2}|-k-mq_1-nq_2\rangle$. Superposition of two orthogonal stripe orders with $q_1$ and $q_2$ gives overall similar results as the checkerboard case.

Several key aspects of our experimental observations (Fig. S\ref{Fig. S7}A \& E) are qualitatively reproduced in both these cases. In the nodal region, a single dominant branch opens up a $d$-wave gap along the underlying Fermi surface (Fig. S\ref{Fig. S7}E-H and the insets). In contrast, multiple branches of comparable spectral weight are seen in the antinodal region, with one branch exhibiting little dispersion in two dimensions and another dispersive branch which shows back-bending at $k_G$ away from $k_F$.

\subsection{Pair-density-wave pseudogap order}
Other density-wave alternatives for the broken-symmetry pseudogap order have not been (but should be) explored. A particularly interesting candidate is the pair density wave order, as an inherent mixture of density-wave and superconducting correlations \cite{stripe:theory:PDW_1st, stripe:theory:PDW_2nd, stripe:theory:ErezStripedSC}. 

\subsection{Simple nematic pseudogap order}
Nematic order, which breaks rotational symmetry but not translational symmetry of the lattice can also potentially explain our key observations in the superconducting state. We have studied a simple form of a nematic distortion (Pomeranchuk type) of the bare Fermi surface that introduces a difference in $t$ for the $x$ and $y$ directions, $t_y=\alpha t_x$ but not in $t^\prime$, $t^{\prime\prime}$, $t^{\prime\prime\prime}$ ($\varepsilon_0$ is adjusted to maintain the Luttinger's volume). $\alpha$ is chosen so that the experimental $k_G$ falls on the distorted Fermi surface and the superconducting gap opens along the distorted Fermi surface (inset of Fig. S\ref{Fig. S8}D). The tight-binding parameters are chosen to be $t_x, t^\prime, t^{\prime\prime}, t^{\prime\prime\prime}, \varepsilon_0$ \& $\alpha=0.22, -0.034315, 0.035977,-0.0071637, -0.24127$ eV \& 1.0989.

Two types of orthogonal domains exist in a tetragonal crystal. ARPES is expected to detect a superposition of signals from both domains which gives rise to an apparent two-band band structure as shown in Fig. S\ref{Fig. S8}A. The onset of superconductivity produces a gap for each band (Fig. S\ref{Fig. S8}B). Consistent with the experiment, two band features, one dispersionless and the other dispersive showing back-bending, are seen in the antinodal region (Fig. S\ref{Fig. S8}C). But two bands can still be seen away from this region (Fig. S\ref{Fig. S8}D), in contrast with what is seen in experiment. However, one should note that an energy-dependent broadening could in principle make two bands appear to be one when they are close enough. Where the two bands are farthest apart in the antinodal region, an increased coherence of the low-energy band at low temperatures could allow two features to be separately resolved.

\subsection{Pitfalls}
Although these mean-field results present a favorable zeroth-order consistency with the experiment, it is important to point out that they also fail to capture other important aspects of the data. Most obviously, in mean-field theory the quasiparticles are exact eigenstates which produce sharp spectral features; this contrasts with the broad features seen in experiment, e.g., cf. Fig. S\ref{Fig. S7}I-J near the $M$ point. We have tried to correct this by including a phenomenological broadening (described below) but this still does not adequately reproduce the data (see the caption of Fig. S\ref{Fig. S7}). These discrepancies probably reflect the effects of finite correlation length \cite{HTSC:PseudogapDispersion} and strong electron correlations.

\subsection{Quasiparticle lifetime broadening introduced to the simulations}
\begin{figure}
\hspace*{2cm}
\includegraphics [width=4.5in]{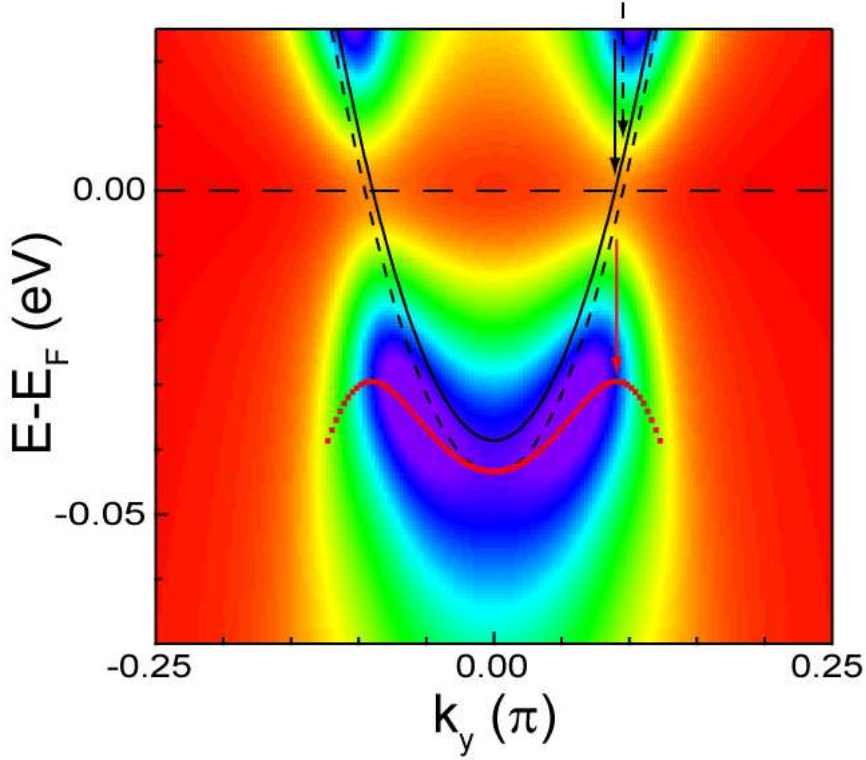}
\caption[ARPES simulation] {\textbf{ARPES simulation on the $k_G-k_F$ misalignment issue.} Even in a Bardeen-Cooper-Schrieffer (BCS) superconductor, {\bf 1)} a finite shift in $k_F$ can result from a finite shift in the chemical potential assuming a particle number conservation in the gapped state. We find that the chemical potential increases by 5 meV in the superconducting state ($\Delta= 35$ meV), which results in an effective bare band at low temperature (dashed black curve) with an increased $k_F$ relative to the bare band in the ungapped state (solid black curve). This increase is too small to be able to account for the $k_G-k_F$ misalignment experimentally observed; {\bf 2)} a finite $k_G-k_F$ misalignment can in principle result from a finite momentum dependence of the superconducting order parameter along the cut. Again this effect alone is negligible in our case, as shown by the simulated false-color plot for the spectral function: the EDC peak dispersion (red dotted curve) shows a back-bending with $k_G$ very close to $k_F$ (of the solid black curve); {\bf 3)} this simulation with realistic experiment conditions (see the \textbf{Simulations} part) also rules out a trivial interpretation of the observed $k_G-k_F$ misalignment due to finite energy, momentum resolutions and quasiparticle lifetime.} 
\label{Fig. S6}
\end{figure}

\begin{figure}
\hspace*{2cm}
\includegraphics [width=4.5in]{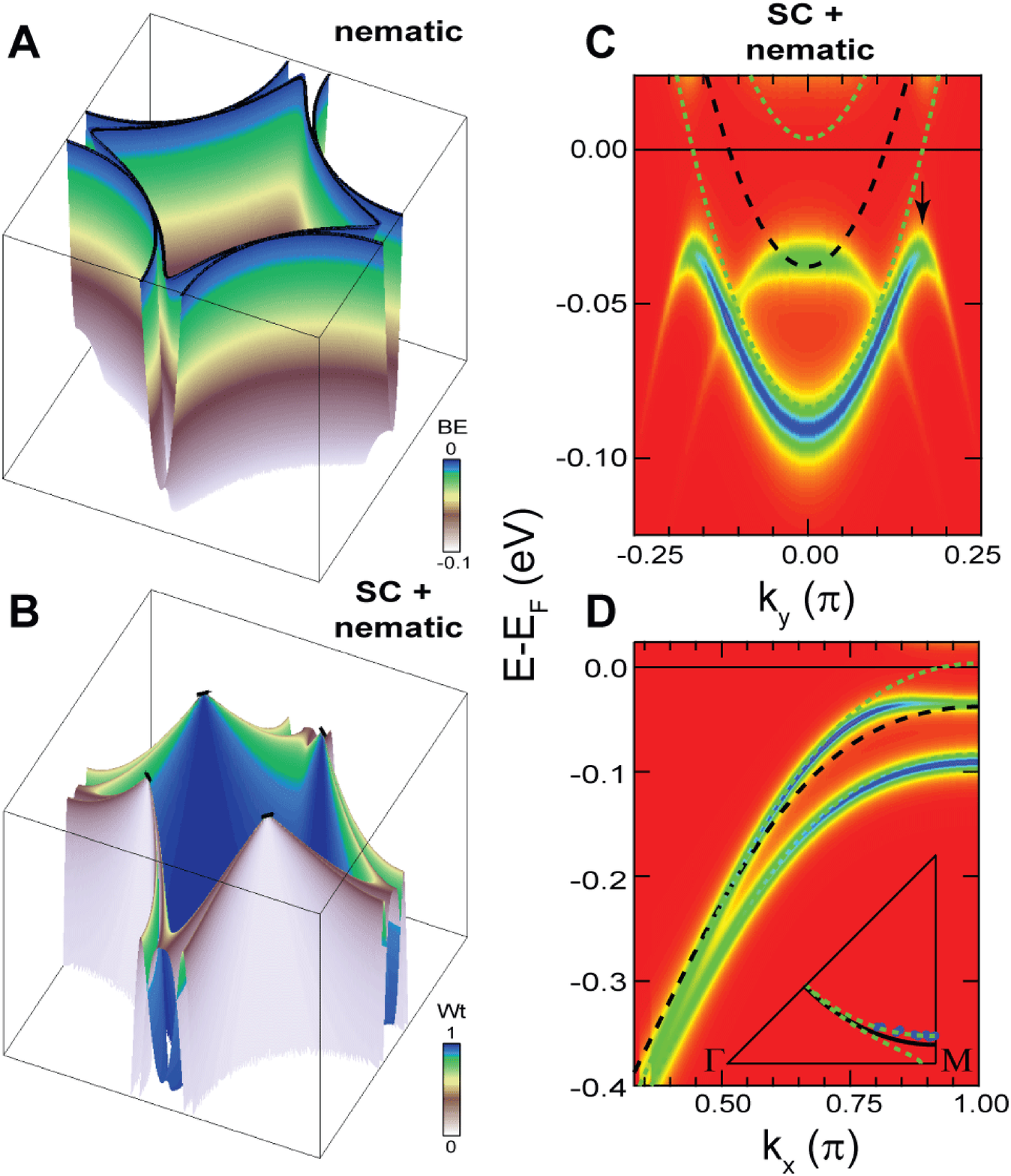}
\caption[Nematic order] {\textbf{Simulations for the superconductivity-nematic-order coexistence.} \textbf{A} (\textbf{B}), Cartoon for the band structure in the nematic phase without (with) coexisting $d$-wave superconductivity ($\Delta= 35$ meV). Two sets of bands from orthogonal domains are superimposed. \textbf{C} (\textbf{D}), Renormalized band dispersions perpendicular to (along) $\Gamma$-M in the coexisting state. Black (green) dashed curves are the bare (distorted) band dispersions. Renormalized band dispersions are moderately broadened for visualization purpose. Back-bending is pointed out by arrow. Insets: $k_G$ from the experiment that falls on the distorted Fermi surface in green. Color scale is proportional to the intrinsic spectral weight in \textbf{B}-\textbf{C}. Note that the values of parameters in the simulation are chosen only for a qualitative agreement with the experiment.} 
\label{Fig. S8}
\end{figure}

All simulation results, except those in Figs. S\ref{Fig. S6} \& S\ref{Fig. S7}K, are shown with the renormalized band dispersion $\varepsilon^\prime(k)$ broadened by a Fermi-liquid-like energy-dependent linewidth $\Gamma=0.1\omega^2+0.005$ (eV) and its intensity proportional to $|u_k|^2$. Fig. S\ref{Fig. S7}K assumes a phenomenological (marginal-Fermi-liquid-like) spectral function with a linewidth, $\Gamma=\sqrt{(\alpha\omega)^2+(\beta T)^2}$ (eV), where $\alpha=$0.715 and $\beta=$2.57. This model was demonstrated to provide a reasonable global fit to the experimental antinodal spectra of heavily-overdoped Bi2212 ($T_c$= 65 K) taken at various photon energies (16.6 - 32 eV) within energy (-200 - 50 meV) and temperature (10 - 128 K) ranges comparable to our study \cite{HTSC:PDH_Bi2212_BBS2}.

\subsection{Effects of finite experimental resolutions and quasiparticle lifetime}
The ARPES simulation in Fig. \ref{Fig. S6} is based on a phenomenological spectral function in the superconducting state \cite{misc:PES:LEGsim}. We use the same global tight-binding bare band structure and assume $d$-wave superconductivity ($\Delta= 35$ meV). The experimental energy and momentum resolutions and a realistic quasiparticle linewidth ($\alpha=$0.715 and $\beta=$2.57) are incorporated. We find that these factors have a negligible effect on the $k_G-k_F$ misalignment. 

\section{Additional discussion}
\subsection{On the nature of the broken-symmetry pseudogap state}

An outstanding task after presenting our data in the main text is to identify the nature of the broken-symmetry state, and to relate it to the apparently similar electronic changes that occur below $T^*$ in other cuprates. There are two real-world considerations which complicate this program: {\bf 1)} To the extent that the pseudogap state involves broken spatial symmetries, the role of quenched disorder is typically severe. It is expected to round the transition and to limit the growth of the pertinent correlations below $T^*$ to a limiting length scale which diverges only as the strength of the disorder tends to zero. This means that macroscopic measures of broken symmetry, especially where (as in Bi2201) the dopant atom distribution produces an inescapable degree of randomness, are expected to be less precise than they are in idealized models. {\bf 2)} In comparing different materials, even if the same ~``universal'' physics underlies the pseudogap, there can be detailed material-specific differences, even in the precise patterns of broken symmetry involved, making cross-material comparisons more subtle than they would be in idealized models.

While nominally the onset of PKE at $T^*$ suggests that the pseudogap phase uniformly breaks time-reversal symmetry, the small size of the effect and the lack of any directly detectable ferromagnetism below $T^*$ make this conclusion uncertain, as previously discussed in the case of YBCO \cite{HTSC:KerrEffect_YBCO}. A recent zero-field nuclear magnetic resonance (NMR) study detected no static magnetism on the Cu sites in superconducting Bi2201 \cite{cuprates:Bi2201:NMR_knightshift_PG}, implying that either the magnetism has a very novel character \cite{HTSC:CompetingOrders_DDW, HTSC:CompetingOrders_VarmaLoops} or that there is no magnetism. In either case, the similarity of the PKE onset at $T^*$ in Bi2201 and YBCO supports the proposition that the pseudogap order has a common character in a broad class of cuprate superconductors.

The large magnitude of the antinodal gap ($\sim 3.5 k_BT^*$, measured relative to $E_F$) at temperatures well below $T^*$ and the fact that it removes a substantial portion of the Fermi surface, implies that the spectral changes detected in ARPES reflect the primary pseudogap order. This is consistent with what has been inferred from the temperature dependence of the Knight shift measured by NMR \cite{cuprates:Bi2201:NMR_knightshift_PG} of bulk Bi2201 in high enough magnetic fields to suppress superconductivity: The density of states (DOS) at $E_F$ begins to drop sharply at $T^*$, exhibiting a similar temperature dependence as those shown in Fig. 3. However, also consistent with the ARPES, a substantial fraction of the DOS remains ungapped down to 0 K. 

\subsection{Implications with various reported candidates for the pseudogap order}

Here we briefly compare the present results with studies of putative pseudogap order in other cuprate superconductors. Strong evidence has been found of nematic order in YBCO
 \cite{stripe:neutron:YBCO_eLiquidCrystal,HTSC:PseudogapBreakRotational} and Bi2212 \cite{stripe:STM:CB_CB2212_Kapitulnik, stripe:STM:Nematic_Bi2212}. Evidence of unconventional translation-symmetry-preserving antiferromagnetism in YBCO \cite{HTSC:BreakTimeRev_neutron_YBCO} and HgBa$_2$CuO$_{4+\delta}$ \cite{HTSC:BreakTimeRev_neutron_Hg1201} has also been reported. These experiments suggest the pseudogap may primarily involve a subtle form of discrete symmetry-breaking with the wave vector $\vec Q = 0$. While such order could produce significant shifts in the electron dispersion, without breaking translational symmetry (or particle conservation), however, it is hard to see, at least at mean-field level, how it would lead to a gap in the spectrum. There is substantial evidence of unidirectional spin and charge density wave order \cite{stripe:neutron:TranquadaStripe1st} in various cuprates, which spontaneously break lattice translational symmetry. Such order could readily lead to the opening of a gap on the nested portions of the Fermi surface. However, direct evidence of the existence of such a state has been found by neutron and X-ray scattering experiments only at temperatures below 50$\sim$ 60 K (well below $T^*$) in particularly ~``stripe-friendly" materials. Unless the order parameter is somehow anomalously difficult to observe in diffraction \cite{HTSC:CompetingOrders_DDW}, it is hard to believe that similar experiments would not have already seen new Bragg peaks in the pseudogap regime of many cuprates.

\subsection{On the ordering vector of a putative density-wave pseudogap order}

Ideally, by analyzing the momentum dependence of the spectral changes in ARPES spectra induced by the pseudogap, one might be able to back out information concerning the nature of the pseudogap order, such as the ordering vector, $\vec Q$, of a putative density-wave state. In mean-field theory, starting from a bare band structure, $\epsilon_{\vec k}$, as the temperature $T$ is reduced below $T^*$, the pseudogap-induced spectral shifts would be expected to grow in proportion to $(T^*-T)^\beta$ with $\beta = 1/2$ only at special points in $\vec k$ space for which $\epsilon_{\vec k} = \epsilon_{\vec k-\vec Q}$. In all other cases such shifts would be proportional to $(T^*-T)^{2\beta}/|\epsilon_{\vec k} - \epsilon_{\vec k-\vec Q}|$. From Fig. 3, to the extent that the spectral shift at $k_F$ and the change of Kerr rotation near $T^*$ can be associated with a critical exponent, they are both proportional to $(T^*-T)^{2\beta}$ with $\beta\approx 1/2$. The former is thus consistent with the mean-field expectation for a poor nesting of the antinodal portion of Fermi surface, which is in turn supported by the observed $k_G-k_F$ misalignment. A nematic order with $\vec Q=0$ expects a scaling of $(T^*-T)^\beta$, which seems not quite consistent with this particular aspect of our data.

Although $\vec Q (\neq 0)$ can in principle be determined by studying the critical exponent as a function of position along the Fermi surface, there is every reason to think that order parameter fluctuations will significantly broaden both the temperature and momentum dependences of the spectroscopic signatures of the transition (even when the transition itself characterized by transport measurements remains sharp). Moreover, although the bare electronic structure is simple, its strange-metal nature precludes a safe description in terms of non-interacting quasiparticles with dispersion $\epsilon_{\vec k}$. 




\clearpage

\begin{FPfigure}
\hspace*{-0.5cm}
\includegraphics [width=6.8in]{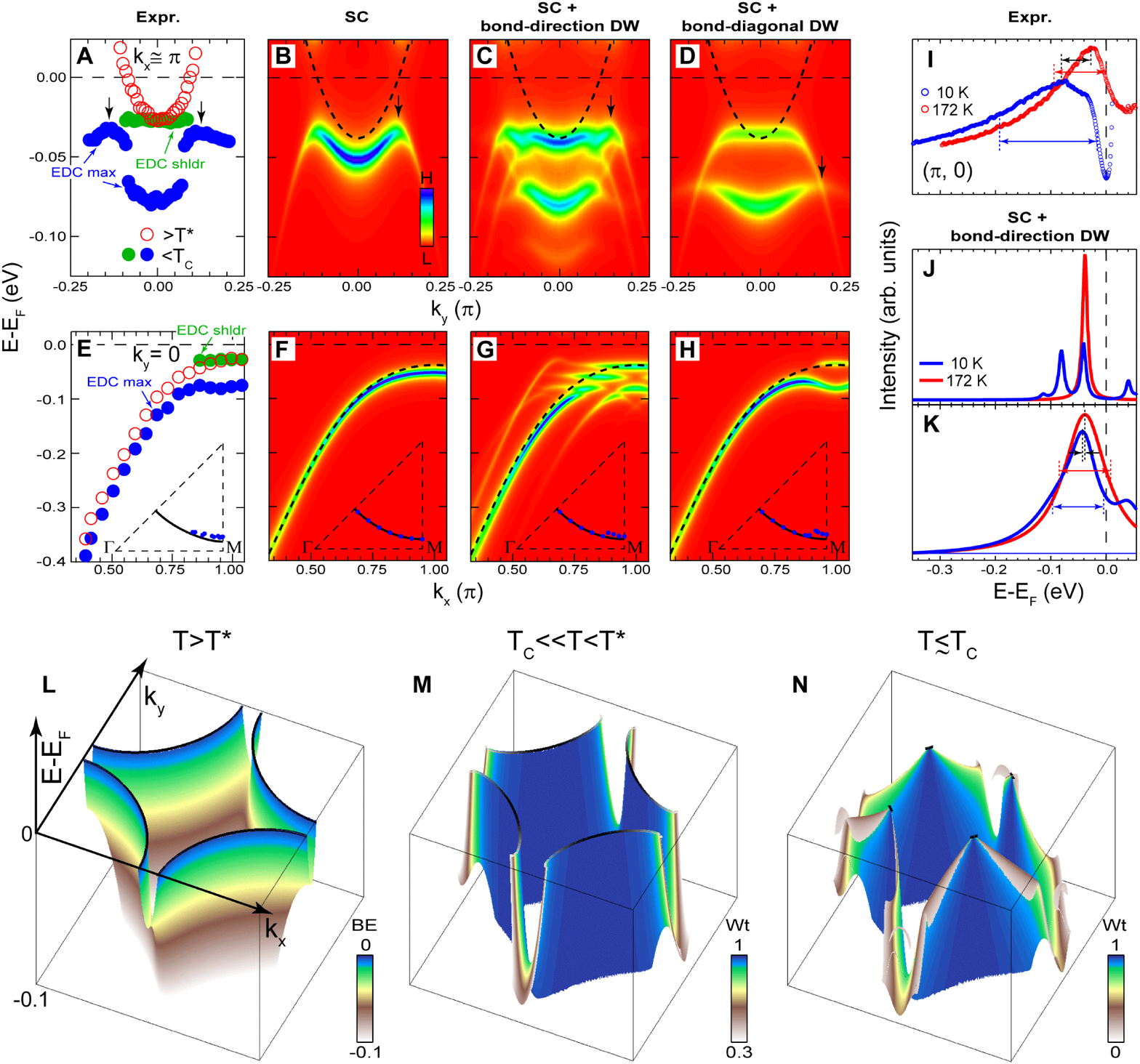}
\caption[Simulations for the superconductivity-density-wave coexistence] {\textbf{Simulations for the superconductivity-density-wave coexistence.} \textbf{A} \& \textbf{E}, Summary of experimental dispersions at 10 K and 172 K along Cut C1 (reproduction of Fig. 2O) and P1-P16 (Fig. 2V-W), respectively. \textbf{B} \& \textbf{F}, \textbf{C} \& \textbf{G}, \textbf{D} \& \textbf{H}, Renormalized band dispersions by simulations assuming different long-range orders: $d$-wave superconductivity (SC, $\Delta= 35$ meV), its coexistence with bond-direction $q_1=(0.15\pi,0)$ \& $q_2=(0, 0.15\pi)$ checkerboard density wave (DW, $V_1= 20$ meV) or with bond-diagonal $q_{AF}=(\pi,\pi)$ density wave ($V_2= 35$ meV), respectively, perpendicular to or along $\Gamma$-M. Dashed curves are bare band dispersions resulting from a global tight-binding fit to the experimental EDC peak dispersions at 172 K (Fig. \ref{Fig. S5}). Renormalized band dispersions are moderately broadened for visualization purpose and the intensity reflects intrinsic spectral weight. Note that the visual separation between dominant states at low and high energies is not a real hybridization gap but due to suppression of intrinsic spectral weight of dispersions within (\textbf{C}-\textbf{D}). Insets of \textbf{E}-\textbf{H}: $k_G$ (arrows in \textbf{A}-\textbf{D}) shown in half of a quadrant, from the experiment and simulations. Apparent discontinuities in the insets of \textbf{G}-\textbf{H} are due to changes of the physical character of back-bending. Error bars are smaller than the symbol size. \textbf{I}-\textbf{K}, Comparison between experimental and simulated EDCs at M point at 10 K and 172 K. Experimental intensity is normalized by the incident photon flux. Simulated EDCs in \textbf{J} are obtained from \textbf{G}, which are broadened phenomenologically in \textbf{K} (see the \textbf{Simulations} part). Different arrows are eyeguides for the EDC linewidth and energy shift of the EDC centroid. In \textbf{K}, multiple features below $E_F$ at low temperature merge into a single feature, with its centroid energy and overall linewidth barely changed from high temperature, still very different from the experiment (\textbf{I}). \textbf{L}-\textbf{N}, Cartoons for the evolving band structure in different phases. Fermi surface is shown in black at $E_F$. Color scale is proportional to the intrinsic spectral weight in \textbf{M} \& \textbf{N}. A $(\pi,\pi)$ density wave order ($V_2= 50$ meV) and its coexistence with $d$-wave superconductivity ($\Delta= 35$) are assumed in \textbf{M} \& \textbf{N}, respectively, only for illustrative purpose. Note that the values of parameters in all simulations are chosen only for a qualitative agreement with the experiment.}
\label{Fig. S7}
\end{FPfigure}

\end{document}